\documentclass[aps,pre,twocolumn,showpacs,10pt,superscriptaddress]{revtex4-2}

\usepackage{graphicx}% Include figure files
\usepackage{dcolumn}% Align table columns on decimal point
\usepackage{bm}% bold math
\usepackage{amsmath}
\usepackage{amssymb}
\usepackage{float}
\usepackage[titletoc,title]{appendix}
\usepackage[ruled,vlined]{algorithm2e}

\newcommand{\expr}[1]{$\langle${#1}$\rangle$}

\begin{document}

%%%%%%%%%%%%%%%%%%%%%%%%%%%%%%%%%%%%%%%%%%%%%%%%%%%%%%%%%%%%%%
\author{Jirawat Tangpanitanon}
\email{jirawat@qtft.org}
\affiliation{Quantum Technology Foundation (Thailand), Bangkok, Thailand}
\affiliation{Thailand Center of Excellence in Physics, Ministry of Higher Education, Science, Research and Innovation, Bangkok, Thailand}
\author{Jirawat Saiphet}
\email{jirawat.sai@outlook.com}
\affiliation{Optical and Quantum Physics Laboratory, Department of Physics, Faculty of Science, Mahidol University, Bangkok, Thailand}
\author{Pantita Palittapongarnpim}
\affiliation{Chula Intelligent and Complex Systems Lab, Department of Physics, Faculty of Science, Chulalongkorn University, Bangkok, Thailand}
\author{Poompong Chaiwongkhot}
\affiliation{Optical and Quantum Physics Laboratory, Department of Physics, Faculty of Science, Mahidol University, Bangkok, Thailand}
\author{Pinn Prugsanapan}
\affiliation{Kasikorn Business-Technology Group, Bangkok, Thailand}
\author{Nuntanut Raksasri}
\affiliation{Kasikorn Business-Technology Group, Bangkok, Thailand}
\author{Yarnvith Raksri}
\affiliation{Kasikorn Business-Technology Group, Bangkok, Thailand}
\author{Pairash Thajchayapong}
\affiliation{Kasikorn Business-Technology Group, Bangkok, Thailand}
\author{Thiparat Chotibut}
\email{thiparat.c@chula.ac.th}
\affiliation{Chula Intelligent and Complex Systems Lab, Department of Physics, Faculty of Science, Chulalongkorn University, Bangkok, Thailand}
%%%%%%%%%%%%%%%%%%%%%%%%%%%%%%%%%%%%%%%%%%%%%%%%%%%%%%%%%%%%%

\title{Hybrid Quantum-Classical Algorithms for Loan Collection Optimization \\with Loan Loss Provisions}

\date{\today}
%%%%%%%%%%%%%%%%%%%%%%%%%%%%%%%%%%%%%%%%%%%%%%%%%%%%%%%%%%%%%

\begin{abstract}
Banks are required to set aside funds in their income statement, known as a loan loss provision (LLP), to account for potential loan defaults and expenses. By treating the LLP as a global constraint, we propose a hybrid quantum-classical algorithm to solve Quadratic Constrained Binary Optimization (QCBO) models for loan collection optimization. The objective is to find a set of optimal loan collection actions that maximizes the expected net profit presented to the bank as well as the financial welfare in the financial network of loanees, while keeping the LLP at its minimum. Our algorithm consists of three parts: a classical divide-and-conquer algorithm to enable a large-scale optimization, a quantum alternating operator ansatz (QAOA) algorithm to maximize the objective function, and a classical sampling algorithm to handle the LLP. We apply the algorithm to a real-world data set with 600 loanees and 5 possible collection actions. The QAOA is performed using up to 35 qubits on a classical computer. We show that the presence of the QAOA can improve the expected net profit by approximately $70\%$, compared to when the QAOA is absent from the hybrid algorithm. Our work illustrates the use of near-term quantum devices to tackle real-world optimization problems.
\end{abstract}

\maketitle
%%%%%%%%%%%%%%%%%%%%%%%%%%%%%%%%%%%%%%%%%%%%%%%%%%%%%%%%%%%%%

\section{Introduction}
Financial institutions often rely on the ability to solve computationally intensive problems. Quantum computing holds the promise of highly efficient algorithms that can provide speedup over some best-known classical algorithms \cite{Orus2019, Egger2020,2020arXiv201106492B}. Applications of quantum computing on finance have been recently explored both theoretically and experimentally on small quantum devices. Examples include portfolio optimization \cite{Hodson2019, Cohen2020, mugel2012, davide2019}, option pricing \cite{nikitas2020, ramo2021, fontanela2019}, risk analysis \cite{braun2021}, transaction settlement \cite{braine2019}, and credit valuation adjustment\cite{alcazar2021}.

In this work, we introduce a new use-case of quantum computing in finance, namely loan collection optimization. The goal is to find a set of optimal actions to be taken on the loanees in order to maximize the expected net profit presented to the lender. Examples of these actions include doing nothing, creating a promise-to-pay agreement, debt restructuring and offering various forms of discounted payoffs (DPOs) \cite{Harding2002}. DPOs typically occur in distressed loan scenarios, where loanees are experiencing financial or operational distress, default, or are under bankruptcy. Usually, DPOs are a last resort for lenders because they often involve taking a loss as the loan is repaid for less than the outstanding balance. 
 
Loan collection actions are normally chosen based on the history of an individual loanee and the experience of the collector. However, the network-induced `domino' effects of such actions, where financial distress can potentially cascade through interconnected financial network of loanees, are usually ignored due to the complexity of the underlying financial system \cite{marco2021}. For example, offering a DPO to one loanee usually signifies the cash flow problem of such loanee which could cause disruptions in the supply chain \cite{nin2021}. Such scenarios would negatively impact cash flow of related parties. Some of which may also be the customer of the lender. Therefore, it is important to account for financial association among the loanees when determining the optimal collection actions. The lender should aim to maximize the expected net profit while minimizing the potential cascade of financial distress on the financial network of the loanees.

Another important aspect of loan collection is Loan Loss Provision (LLP) \cite{OZILI2017144}. LLP is regulated by the government to ensure the financial health of the bank. The idea is to set aside funds as an expense in the financial statement to account for potential loan defaults and expenses  that occur as a result of lending. The amount of LLPs depends on various factors such as types of loanees (individuals, small businesses, large corporations), types of loans, late payments, collection expenses, as well as the collection actions that will be taken. LLPs typically constitute a significant portion of the financial statement. For example, in the first half of 2020, seventy banks worldwide report under Expected Credit Loss accounting provisions totaling in $\$$161 billion \cite{Kiarelly2021}. Therefore, it is crucial to choose optimal collection actions that keep LLPs at a minimum to promote financial liquidity.

Here, we model the above problem including the financial network and LLPs as Quadratic Constrained Binary Optimization (QCBO) problems with both global and local constraints. We solve this class of problems by devising a hybrid quantum-classical algorithm which consists of three parts: a classical divide-and-conquer algorithm to enable a large-scale optimization, a quantum alternating operator ansatz (QAOA) algorithm to maximize the objective function, and a classical sampling algorithm to handle the LLP. By benchmarking with a real-world data set, we show that the presence of the QAOA can improve the expected net profit by approximately $70\%$, compared to when the QAOA is absent from the hybrid algorithm. Our work can be implemented on near-term quantum devices, consisting of a few tens of qubits and sparse qubit connectivity.

%%%%%%%%%%%%%%%%%%%%%%%%%%%%%%%%%%%%%%%%%%%%%
\section{problem formulation}
\label{sec:problem}

Loan collection is a management problem that involves identifying which collection actions to be taken to which loanees at what time. Due to the complexity of the process, it is a common practice to follow a rule-based guideline for collection activities. However, the actions and time can also be personalized, as is the recent approach to loan collection \cite{Shoghi2019, Liu2019}, in order to respond to the loanee's unique financial history. Moreover, as discussed earlier, one collection action taken to a loanee can affect the financial welfare of other loanees in the financial network which, in turn, affects the expected net profit of the lender. 

To capture both the personalized loan collection approach and the effect of loanee's financial network, we propose a heuristic QCBO model that captures the impact of loan collection actions at a given time for a given loan product. The objective (yield) function is defined as
\begin{equation}
Y = (1-\epsilon)\sum\limits_{i=1}^{N}\sum\limits_{j=1}^{M} h_{i,j}x_{i,j} +\epsilon\sum\limits_{\langle i,i' \rangle}^NA_{i,i'}(1-x_{i,1})(1-x_{i',1}),
\label{eq:objective_function}
\end{equation}
where $N$ is the number of loanees and $M$ is the number of possible actions. Here, $\{x_{i,j}\}$ are binary decision variables: $x_{i,j}=1$ if action $j$ is taken to loanee $i$ and $x_{i,j}=0$ otherwise. By convention, we set $j=1$ to denote the DPO action. The coefficient $h_{i,j}\in\mathbb{R}$ is an expected net profit that the bank would get if action $j$ is taken to loanee $i$ \footnote{The coefficient can also be seen as a way of ranking the bank's preferred action towards a loanee and can be personalized to the loanee's loan payment history. One approach for the personalized calculation of $h_{i,j}$ is to model the loan collection problem as a Markov Decision Process (MDP) \cite{Shoghi2019}.}. $(A_{i,i'})\in\mathbb{R}^{N\times N}$ is an association matrix, defined as the averaged transaction between loanees $i$ and $i'$. According to the form of $Y$, association or `cash flow' between loanees $i$ and $i'$ vanishes if one of them is offered a DPO, i.e. $x_{i,1}=1$ or $x_{i',1}=1$. The hyper-parameter $\epsilon\in[0,1)$ tunes the competition between the expected return to the bank and  the financial welfare in the network of loanees. Increasing $\epsilon$ will cause the optimal solution to contain a fewer number of DPO actions being taken, which increases the overall financial welfare of the loanees.      

We now introduce local constraints that force only one action to be taken to each loanee as
\begin{equation}
\sum\limits_{j=1}^{M}x_{i,j} = 1, 	\forall i\in\{1,...,N\}.
\label{eq:local_constraints}
\end{equation}
LLP is treated as a global constraint as 
\begin{equation}
\sum\limits_{i=1}^{N}\sum\limits_{j=1}^{M} l_{i,j}x_{i,j} \leq L,
\label{eq:global_constraints}
\end{equation}
where $L\in\mathbb{R}^+$ is the upper bound for the total provision and $l_{i,j}\in\mathbb{R}^+$ is the provision for taking action $j$ to loanee $i$. Determining the values of $l_{i,j}$ depends on probability of default estimates and expected loss. Exact details are bank-specific and depends on government regulation, such as the Third Basel Accord, an international, voluntary regulatory framework for banks developed by the Basel Committee on Banking Supervision \cite{OZILI2017144}. 

Although $L$ can be considered a fixed value and a hard constraint, in practice when we are predicting the net income of a known collection of loans, we might allow some leniency in terms of $L$ in favour of a solution that yields high net income. In this scenario, we treat Eq. (\ref{eq:global_constraints}) as a soft constraint with a secondary objective of minimizing the total provision during optimization.

%%%%%%%%%%%%%%%%%%%%%%%%%%%%%%%%%%%%%%%%%%%%%%%%%%%%%%%%%%%%%%

\begin{figure*}
\center
\includegraphics[width=0.95\textwidth]{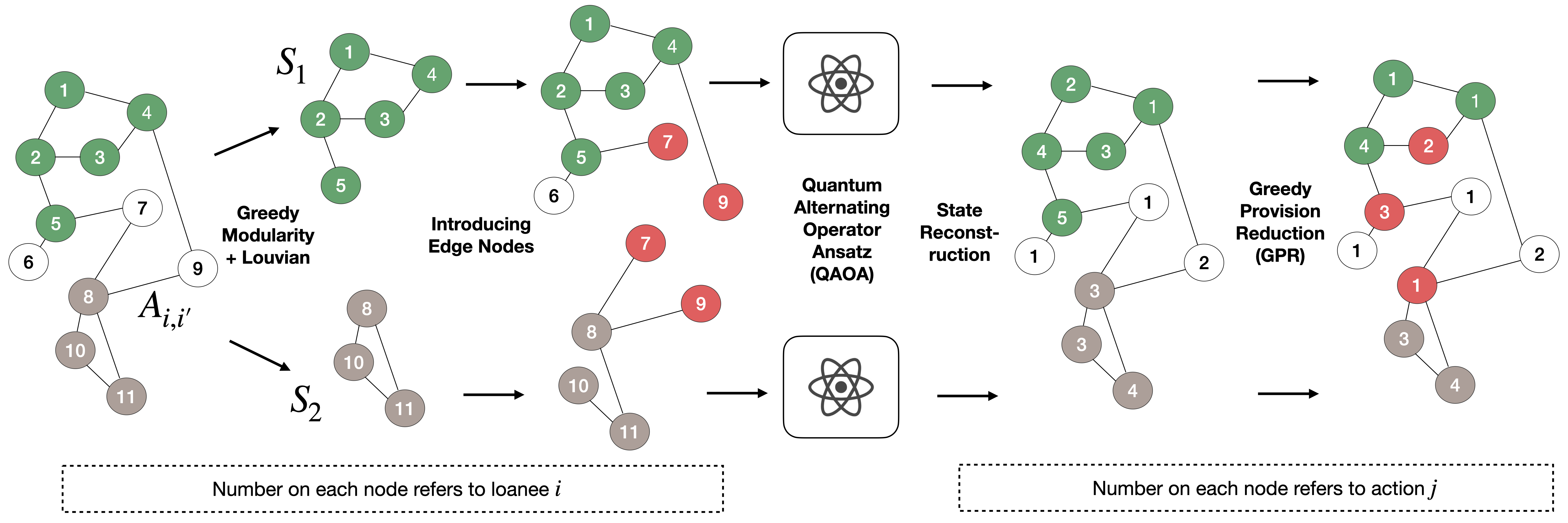}
\caption{\textbf{Schematic of the hybrid quantum-classical algorithm.} Firstly, the financial transaction network as indicated by $A_{i,i'}$ is divided into small groups using the Greedy Modularity algorithm and the Louvian method. The edge nodes, labelled in red, are introduced to each group to accommodate common loanees among the two groups. The optimal set of collection actions for each group is obtained via the QAOA. Solutions from different groups are then combined to reconstruct the optimal actions for the original problem. The latter is fed to the GPR algorithm to minimize LLP. See the main text for more details.}
\label{fig1}
\end{figure*}

\section{Hybrid Quantum-Classical Algorithm }

In this section, we present our algorithms, as depicted in Fig. \ref{fig1}, to find a set of optimal actions $\{x_{i,j}\}$ that maximizes $Y$ subjected to the constraints in Eq. (\ref{eq:local_constraints}) and Eq. (\ref{eq:global_constraints}). The idea is to use a classical divide-and-conquer algorithm \cite{li2021} to arrange the loanees into small groups based on the association matrix. Loanees from the same group will have a higher association compared to those outside the group. Next, we find the optimal collection actions for each group using the Quantum Alternating Operator Ansatz (QAOA) algorithm \cite{a12020034}, without considering LLPs. Actions from different groups are then combined to reconstruct the optimal actions for the original problem. The latter process is referred to as state reconstruction. Lastly, we apply a classical sampling method which we call Greedy Provision Reduction (GPR) to adjust some actions to minimize LLPs while keeping the negative impacts of those adjustments on $Y$ at a minimum. We lay out the details of each step below. The pseudo-code is provided in Appendix \ref{appendix:pseudocodes}.

\textit{Division algorithm -.} To divide the loanees into groups, we rely on the use of two standard community detection algorithms, namely the Clauset-Newman-Moore greedy modularity maximization \cite{PhysRevE.70.066111} and the Louvain method \cite{PhysRevE.80.056117}. After the first iteration, each loanee will be assigned to exactly one group. For example, in Fig. \ref{fig1}, the two groups $S_1$ and $S_2$ contain loanees $i=1,2,3,4,5$ and $i=8,10,11$, respectively \footnote{We note that, in general, the combination of both community detection algorithms will iterate until modularity maximization is convergence, which can result in multiple group partition, as opposed to the example schematic in Fig. \ref{fig1}}. As each group will first be optimized separately, the sparse connection between $S_1$ and $S_2$ through loanees $i=7,9$ will be ignored. To mitigate this algorithmic artefact in this divide-and-conquer approach, we introduce `edge nodes', which are outsider loanees that have an association with $S_1$ and $S_2$. As shown in Fig. \ref{fig1}, the groups $S_1$, $S_2$ have edge nodes $i=6,7,9$ and $i=7,9$, respectively. By adding the edge nodes into each group, the new groups now have loanees $i=7,9$ in common. The latter will play a crucial role during state reconstruction. The Louvian method and the edge node introduction are then applied recursively to $S_1$ and $S_2$ until the size of each group is no greater than some fixed threshold $\nu\in \mathbb{Z}^+$. The latter ensures that the size of each group will be small enough to be run on near-term quantum devices.

\begin{figure}
\includegraphics[width=0.95\columnwidth]{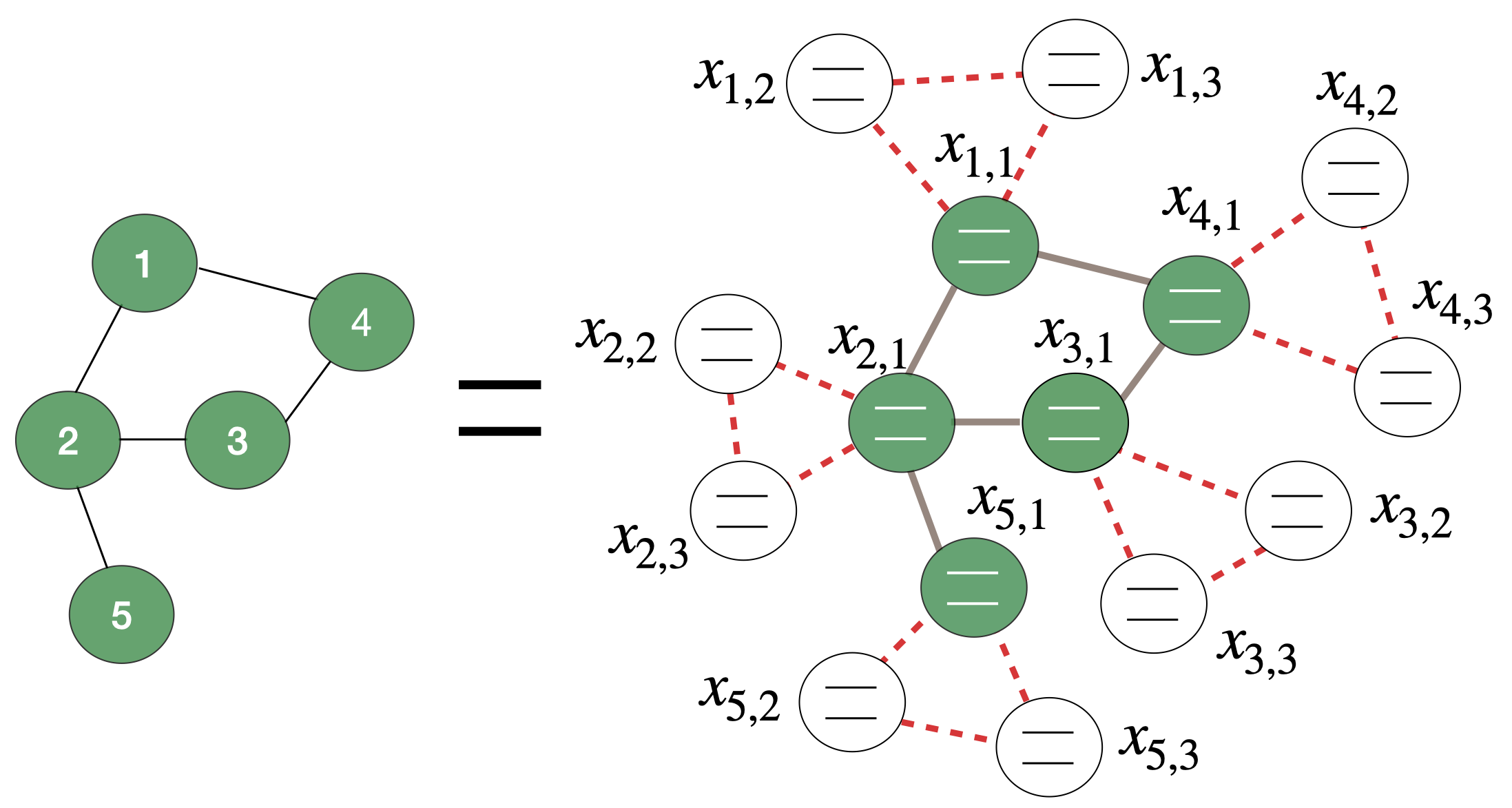}
\caption{\textbf{Qubit representation of loanees.} (Left) A group of loanees and their interactions as indicated by $A_{i,i'}$.(Right) The qubit topology to run the QAOA to find optimal collection actions for this set of loanees. Each loanee is represented by $M$ qubits ($M=3$ in this figure). The interactions among qubits as labelled by solid grey lines and dashed red lines are captured by $\hat{H}_A$ and $\hat{H}_B$, respectively.}
\label{fig2}
\end{figure}

\textit{QAOA algorithm -.} To apply the QAOA, we represent a set of actions applied to the loanees as a basis state of a quantum system consisting of up to $\nu\times M$ qubits. Action $j$ is taken to loanee $i$ ($x_{i,j}=1$) if qubit $(i,j)$ is in the excited state $|1\rangle$; otherwise, the qubit is in the ground state $|0\rangle$. The QAOA is executed by running the following evolution on a quantum hardware,
\begin{equation}
|\psi_T(\underline{\theta})\rangle	= \prod_{t=1}^{T}e^{-i\hat{H}_A\gamma_t}e^{-i\hat{H}_B\beta_t}|\psi_0\rangle,
\end{equation}
where $T$ is the number of driving cycles, $\gamma_t,\beta_t\in \mathbb{R}$ are variational parameters and $\underline{\theta}=\{\gamma_1,\beta_1,..,\gamma_T,\beta_T\}$. The Hamiltonian $\hat{H}_A$ involves interactions among qubits with $j=1$ as labeled by solid grey lines in Fig. \ref{fig2}. It is defined as 
\begin{equation}
	\hat{H}_A = -(1-\epsilon)\sum\limits_{i=1}^{N'}\sum\limits_{j=1}^{M} h_{i,j}\hat{n}_{i,j} -\epsilon\sum\limits_{\langle i,i' \rangle}^{N'}A_{i,i'}(1-\hat{n}_{i,1})(1-\hat{n}_{i',1}).
\end{equation}
Here, $N'\le \nu$ is the number of loanees in the group and $\hat{n}_{i,j}$ is the number operator acting on qubit $(i,j)$. The mixing Hamiltonian $\hat{H}_B$ involves ring-type interactions among qubits with the same $i$ as labeled by red dotted lines in Fig. \ref{fig2}. It is defined as
\begin{equation}
\hat{H}_B = - \frac{J}{2} \sum_{i=1}^{N'}\left[\sum_{j=1}^{M}\left(\hat{X}_{i,j}\hat{X}_{i,j+1}+ \hat{Y}_{i,j+1}\hat{Y}_{i,j}\right)\right],	
\end{equation}
where $\hat{X}_{i,j}$ and $\hat{Y}_{i,j}$ are Pauli's operators acting on qubit $(i,j)$ and $J=1$ is the coupling strength. The periodic boundary condition is applied. 

The initial state $|\psi_0\rangle$ is prepared as a product state such that action $j=1$ is taken to every loanee, i.e., $x_{i,j}=\delta_{j,1}$ where $\delta_{j,1}$ is the Kronecker delta. Note that since both $\hat{H}_A$ and $\hat{H}_B$ preserve the number of qubit excitations per loanee, it follows that all basis states involved in $|\psi_T(\underline{\theta})\rangle$ always satisfies the constraints in Eq. (\ref{eq:local_constraints}).

Following the standard QAOA procedure, we then readout the observable $\langle \hat{H}_A\rangle_{\underline{\theta}}$ from the quantum hardware. The procedure is repeated with a new value of $\underline{\theta}$ as suggested by a classical optimizer until $\langle \hat{H}_A\rangle_{\underline{\theta}}$ is minimized with $\underline{\theta}=\underline{\theta}_{\rm opt}$. The optimal quantum state $|\psi_T(\underline{\theta}_{\rm opt})\rangle$ now concentrates around bit-strings with high objective function values. The probability of sampling a bit string $\underline{x}=\left[x_{1,1},x_{1,2},..,x_{N,M}\right]$ from $|\psi(\underline{\theta}_{\rm opt})\rangle$ is given by $p(\underline{x})=|\langle \underline{x}|\psi_T(\underline{\theta}_{\rm opt})\rangle|^2$, where $|\underline{x}\rangle$ is a basis state representing a set of actions $\underline{x}$.

\textit{Reconstruction algorithm -.} To combine optimal actions from two groups, for example $S_1$ and $S_2$, we simply sample bit-strings from $p(\underline{x})$ of each group. If the bit-strings from $S_1$ and $S_2$ have the same values at the common edge nodes, we append the two bit-strings. For example, as depicted in Fig. \ref{fig1}, $S_1$ has optimal actions $j=2,4,3,1,5,1,1,2$ for loanees $i=1,2,3,4,5,6,7,9$, respectively and $S_2$ has optimal actions $j=1,3,2,3,4$ for loanees $i=7,8,9,10,11$, respectively. Since the common loanees $i=7,9$ have the same actions $j=1,2$ in both $S_1$ and $S_2$. The optimal actions for the combined group $S_1 \cup S_2$ is $j=2,4,3,1,5,1,1,3,2,3,4$ for loanees $i=1,2,...,11$, respectively. If the actions to the common loanees from the two groups are incompatible, then the bit-strings are resampled from $p(\underline{x})$ until the compatible one is found. This procedure is repeated for every group until the optimal actions for the original problem with $N$ loanees is obtained.

\textit{Greedy Provision Reduction (GPR) algorithm -.} Lastly, we apply a classical sampling algorithm to handle the LLP constraint in Eq. (\ref{eq:global_constraints}) by minimizing the total provision. The idea is to quantify the impact of changing action $j$ of loanee $i$ to $j'$ using a finesse score $f_{i,j\to j'}\in \mathbb{R}$. For a given set of actions $\{i,j\}$, there are $N\times M$ values of $f_{i,j\to j'}$'s. We then choose $f_{i,j\to j'}$ with the highest value and update the action accordingly. To define $f_{i,j\to j'}$, we first define a reward $a_{i,j\to j'}$ as a reduction in the total provision and a penalty $b_{i,j\to j'}$ as a reduction in $Y$, when changing action $j$ of loanee $i$ to action $j'$. By definition, the finesse score increases when switching actions result in a higher reward and a lower penalty. Specifically, if the total provision does not increase ($a_{i,j\to j'}>=0$) and $Y$ does not decrease ($b_{i,j\to j'}<=0$), we will choose the action with the highest reward, i.e., $f_{i,j\to j'}=a_{i,j\to j'}$. If both the provision and $Y$ decreases ($a_{i,j\to j'}>0$ and $b_{i,j\to j'}>0$), we set $f_{i,j\to j'}=-b_{i,j\to j'}/a_{i,j\to j'}$. Finally, if the total provision increases ($a_{i,j\to j'}<0$) or the total provision stays constant ($a_{i,j\to j'}=0$) but $Y$ decreases ($b_{i,j\to j'}>0$), we set $f_{i,j\to j'}=-\infty$ indicating that this action will not be chosen. The update procedure is carried out iteratively, one action adjustment at a time, until the total provision is no greater than $L$ as indicated in Eq. (\ref{eq:global_constraints}). In the case $\epsilon=0$, there exists a theoretical bound to ensure that the optimal $Y$ obtained from this procedure will not be less than half of the global optimal value \cite{jon2013}.

%%%%%%%%%%%%%%%%%%%%%%%%%%%%%%%%%%%%%%%%%%%%%%%%%%%%%%%%%%%%%%
\begin{figure}
\includegraphics[width=0.95\columnwidth]{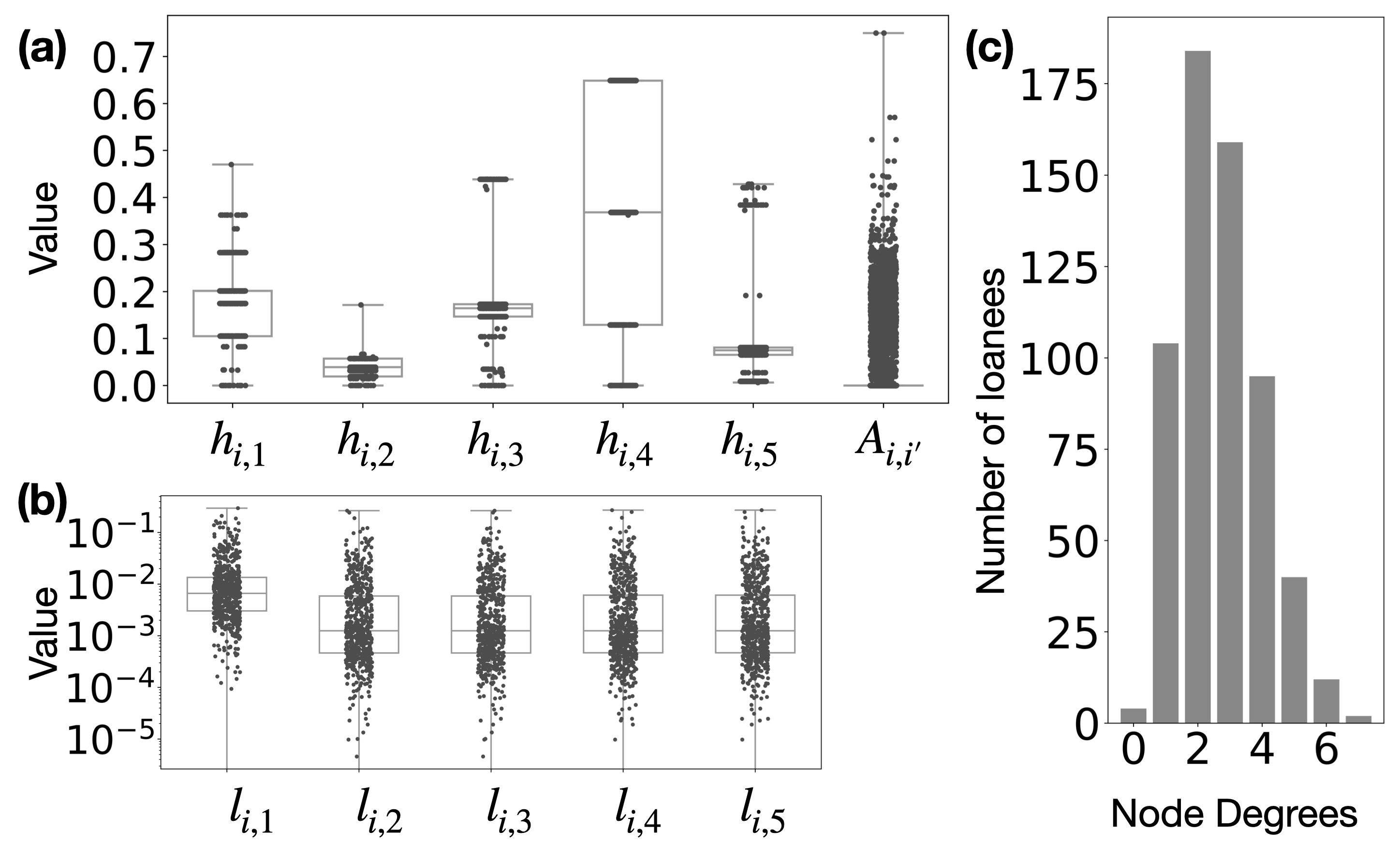}
\caption{\textbf{Statistics of model parameters obtained from Kasikornbank data set.} (a) Box plot of $h_{i,j}$ and $A_{i,i'}$ showing their distributions and quantiles. (b) Box plot of $l_{i,j}$ in a log scale. (c) The histogram of the node degrees in $A_{i,i'}$. }
\label{fig3}
\end{figure}

\begin{figure*}
\center
\includegraphics[width=0.95\textwidth]{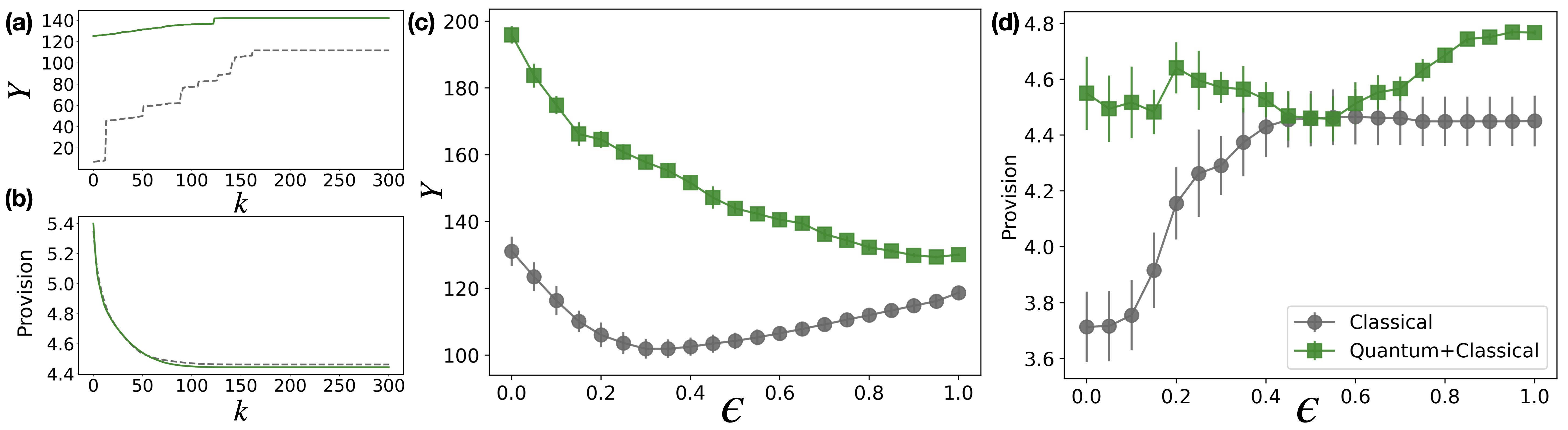}
\caption{\textbf{Performance of the hybrid algorithm}. (a)-(b) shows $Y$ and the total provision during the GPR algorithms with $\epsilon=0.5$ as a function of the number of steps $k$ in GPR, respectively. The solid green lines are when the solutions reconstructed from the QAOA are used as an input and the dashed grey lines are when random solutions are used as an input. (c)-(d) shows $Y$ and the total provision as a function of $\epsilon$, respectively. The square green dots are from the hybrid algorithm and the circle grey dots are from the standalone GPR. ($\nu=7$, $T=2$).}
\label{fig4}
\end{figure*}

\begin{figure}
\includegraphics[width=0.95\columnwidth]{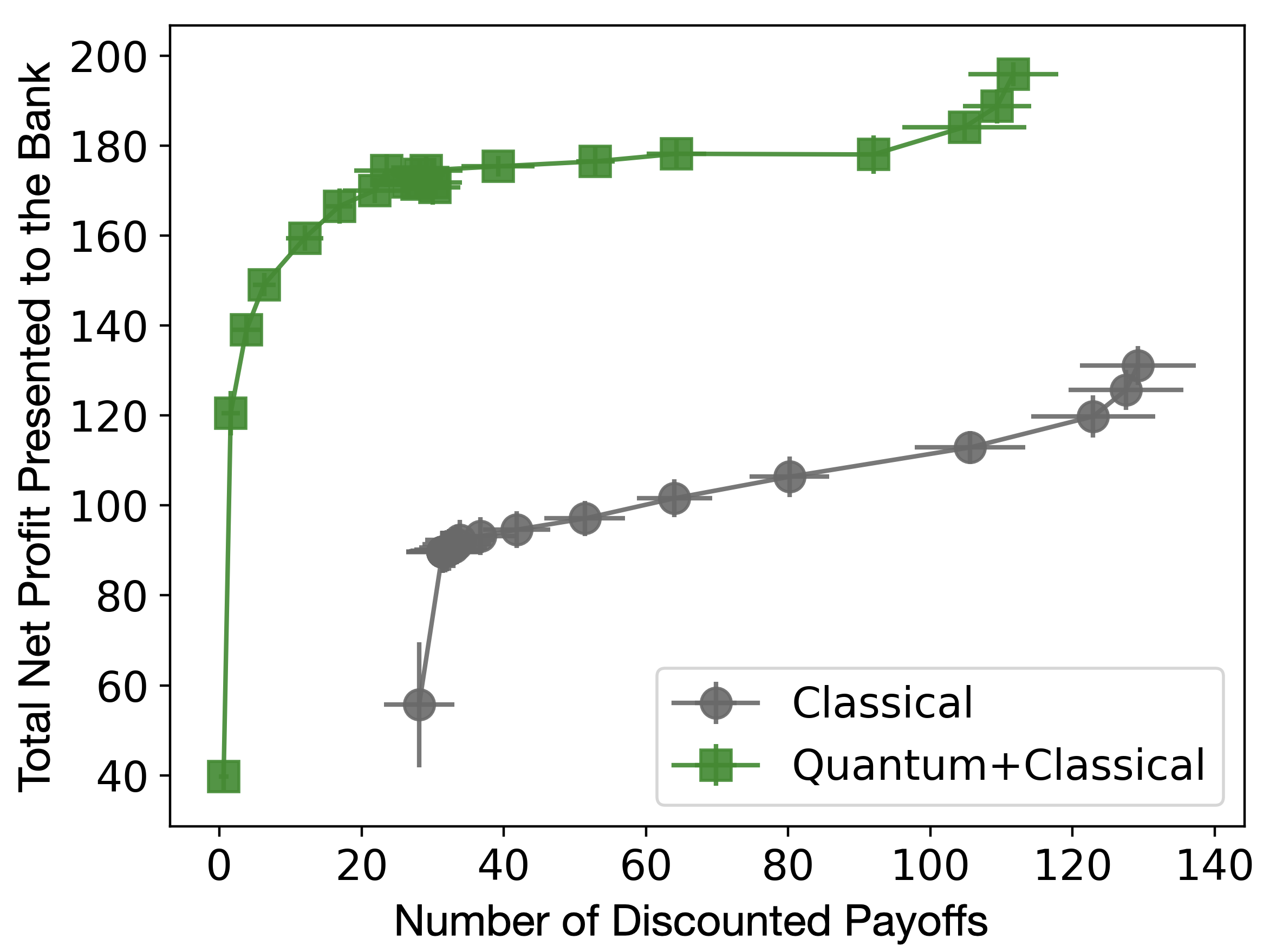}
\caption{Total net profit presented to the bank as a function of the number of DPO actions being taken. The square green dots are from the hybrid algorithm and the circle grey dots are from the standalone classical GPR. Note that our hybrid quantum-classical always yield a better total net profit presented to the bank compared to the standalone classical GPR algorithm ($\nu=7$, $T=2$).}
\label{fig5}
\end{figure}

\section{Results and Discussions}

\subsection{data set}

We use the data set of a personal loan product provided by Kasikornbank. The data set consists of 600 loanees and $5$ possible actions per loanee. $h_{i,j}$ is estimated from $\it{Q}-$learning using the historical data of the loanees \cite{Sutton1998}. $l_{i,j}$ is  related to historical data on repayments and default, loan collection expenses, credit losses, economic conditions, interest rate and tax policy \cite{OZILI2017144}. Specific details of the derivations are omitted here due to data privacy. $A_{i,i'}$ is estimated from transactions between two loanees that happen internally within Kasikornbank. Note that the actual transactions $\tilde{A}_{i,i'}$between the two loanees could be higher than this value, i.e., $\tilde{A}_{i,i'}=\alpha_{i,i'}A_{i,i'}$ with $\alpha_{i,i'}\ge1$. However, it is not possible for a bank to collect every transaction from a customer. Therefore, we assume that $\alpha_{i,i'}$ does not depend on $i$ and $i'$, so that it can be absorbed into the redefinition of $\epsilon$. The interpretation of $\epsilon$ in practice will be discussed later in the text.

The distributions of $h_{i,j}$, $l_{i,j}$, and $A_{i,i'}$ are depicted in Fig. \ref{fig3}. The units are made arbitrary for privacy concerns. The value of $h_{i,j}$ and $A_{i,i'}$ range from $0$ to $0.7$, while the value of $l_{i,j}$ ranges from $\sim 10^{-5}$ to $\sim 1$. The financial network, as indicated by $A_{i,i'}$, is sparse with node degrees ranging from $0$ to $7$. We also benchmark our analysis with a simulated data set in Appendix \ref{appendix:simulateddata} to ensure the generality of our results.

\subsection{Numerical Results}
To analyze the behavior of the QAOA, we will compare our hybrid algorithm with a purely classical algorithm. The latter is achieved by directly feeding a random set of actions to the GPR algorithm. In Fig. \ref{fig4}(a), with $\epsilon=0.5$ where the optimization objective values the expected return to the bank and the financial welfare among loanees equally, we see that the set of actions reconstructed from the QAOA starts with $Y\sim120$ and then increases to $\sim 140$ during the GPR algorithm. On the other hand, a random set of actions start with $Y\sim 10$ and then increases to $\sim 100$, lower than the one with the QAOA. Fig. \ref{fig4}(b) shows that, in both cases, the provision reduces from $\sim 5.4$ to $\sim 4.4$, resulting in $18.5\%$ reduction. We found that for other values of $\epsilon$, $Y$ always monotonically increases, and the provision always monotonically decreases during the GPR algorithm. Hence, there is no need to introduce $L$ to truncate the process. The minimum LLP is treated as a suggested LLP rather than a hard constraint.

Fig. \ref{fig4}(c) shows $Y$ after the GPR algorithm as a function of $\epsilon$. We can see that the hybrid algorithm provides a higher $Y$ compared to the standalone GPR algorithm for all values of $\epsilon$. Fig. \ref{fig4}(d) shows the provision after the GPR algorithm as a function of $\epsilon$. At $\epsilon\lesssim 0.1$, the standalone GPR gives around $17\%$ lower provision compared to the hybrid algorithm. This number is reduced to $\sim 4.3\%$ as $\epsilon$ goes towards unity. This result is expected as the QAOA attempts to increase $Y$ alone without considering the LLPs. Nevertheless, the provision obtained from the hybrid algorithm is kept at $\sim 4.6$ for all values of $\epsilon$. This value is $57.4 \%$ lower than the highest possible provision.

Finally, we note that $\epsilon$ may be hard to interpret in practice because it is normalized by the unknown variable $\alpha_{i,i'}$ as discussed above. To circumvent this, in Fig. \ref{fig5}, we plot the total net profit presented to the bank as a function of the optimal number of DPO actions $N_f$, i.e. the number of loanees with action $j=1$. The latter is, in turn, varied by $\epsilon$. Fig. \ref{fig5} provides an intuitive interpretation of the optimization results. The collector can choose the number of DPO actions based on his/her experience, then read out the estimated net profit that will be made. We found that, for the same $N_f$, the hybrid algorithm always gives a higher expected return. Specifically, for $N_f\sim 30-100$, the expected profit from the hybrid algorithm is $\sim 70\%$ higher than the standalone GPR. In addition, with the hybrid algorithm, we find that the expected return shows a plateau at $N_f\sim 20-100$. This implies that the collector may choose to apply, say, 30 DPOs to get the profit of $\sim 175$, instead of 120 DPOs ($300\%$ more) to get a profit of $\sim 200$ (only $14.3\%$ more).   

%%%%%%%%%%%%%%%%%%%%%%%%%%%%%%%%%%%%%%%%%%%%%%%%%%%%%%%%%%%%%%
\section{Conclusions}

We devise a hybrid quantum-classical algorithm to solve loan collection problems with LLPs by formulating them as QCBO models. Our approach allows the lender to maximize the expected net return while accounting for LLPs and the financial well-being of the loanees, as measured by the association matrix. Compared to a purely classical approach, our hybrid algorithm provides a higher net profit, regardless of the amount of the network effect as measured by $\epsilon$. The algorithm also suggests the collector with the lowest number of DPO actions that still gives a relatively high expected return. Our work paves a way to explore quantum advantage with near-term quantum devices in the financial sector.

%%%%%%%%%%%%%%%%%%%%%%%%%%%%%%%%%%%%%%%%%%%%%%%%%%%%%%%%%%%%%%

\section{Data Availability}
The data that support the findings of this study are available from the corresponding author upon reasonable request.

 \section{Acknowledgement}
We thank Vorapong Suppakitpaisarn for fruitful discussions. P. Palittapongarnpim , P. Chaiwongkhot, and T. Chotibut are supported by the Program Management Unit for Human Resources and Institutional Development, Research and Innovation (Grant No. B05F630108). J.Tangpanitanon, P. Prugsanapan, N. Raksasri and Y. Raksri are supported by Kasikorn Business-Technology Group, Bangkok, Thailand.

\bibliography{ref}
%%%%%%%%%%%%%%%%%%%%%%%%%%%%%%%%%%%%%%%%%%%%%%%%%%%%%%%%%%%%%%%

\pagebreak

\begin{widetext}
\appendix

%%%%%%%%%%%%%%%%%%%%%%%%%%%

%%%%%%%%%%%%%%%%%%%%%%%%%%%%%%%%%

\section{Benchmarking with a Simulated Data Set}
\label{appendix:simulateddata}

In this section, we apply the algorithms presented in the main text to a simulated data set to show the generality of our results. We use $M=5$ and $N=600$ as in the main text. We generate $h_{i,j}$ by sampling from the distribution $ P(i,j)\equiv \left(\log_{10}(j+2) - (1-r_i)\log_{10}(j+1)\right)/\mathcal{C}$, which is a modified Benford's law. Here, $\mathcal{C}$ is a normalization constant, $r_i$ is a random number drawn from the Poisson distribution with the mean value 0.7. This distribution has two peaks: one at $j=1$ (DPO action) and $j=5$ (non-DPO action). The association $A_{i,i'}$ is generated from an Erdos-Renyi ensemble with $N$ nodes and mean degree 2, which has an average path length of approximately $\log (N)/\log (2)$. In Fig. \ref{fig6}, we plot the total net profit present to the bank as the number of DPO actions using optimum solutions from the hybrid quantum-classical and the standalone classical algorithm as in the main text. We found that the hybrid algorithm gives a higher return compared to the classical part as expected. 

\begin{figure}[h]
\includegraphics[width=0.5\columnwidth]{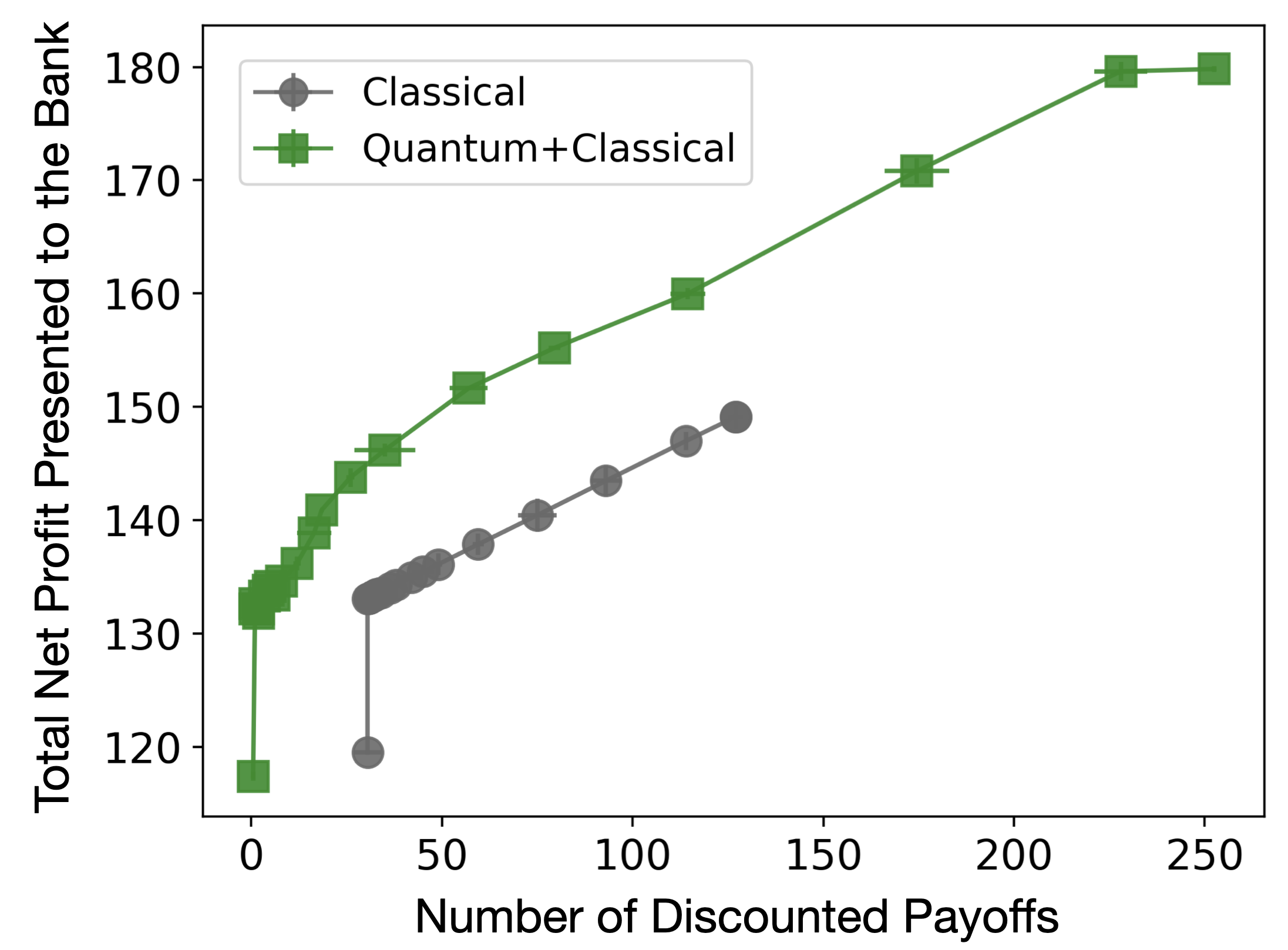}
\caption{Total net profit presented to the bank as a function of the number of DPO actions being taken. The square green dots are from the hybrid algorithm and the circle grey dots are from the standalone classical GPR. ($\nu=7$, $T=2$).}
\label{fig6}
\end{figure}

\section{Pseudo-codes for Hybrid Quantum-Classical Algorithms}
\label{appendix:pseudocodes}

In this section, we provide below pseudo-codes for four algorithms involved in the hybrid quantum-classical algorithms. These are (i) recursive division in the divide-and-conquer algorithm, (ii) QAOA with local constraints, (iii) state reconstruction which is the second part of the divide-and-conquer algorithm, and (iv) Greedy Provision Reduction. See the main text for an overview discussion for each algorithm.

\begin{algorithm}[h]
	\SetKwBlock{Block}{function}{}
	\SetKwInOut{Input}{input}
	\SetKwInOut{Output}{output}
	\SetKw{Return}{return}
	\SetKwFunction{RecLouvain}{RecLouvain}
	
	\Input
	{(i) Association matrix $A=(A_{i,i'})$. \newline (ii) Maximum number of nodes per subgraphs $\nu \in \mathbb{Z}^+$.}
	
	\Output{A set of subgraphs. Each subgraph represents a group of loanees with high association among them. Each node in each subgraph represents one loanee. }
	
	\Begin
	{
		$\mathcal{S} \gets \{\}$  \\
		$G \gets$ \expr{a graph obtained from $A$ } \\
		$\mathcal{S}^{(1)} \gets$ \expr{a set of subgraphs of $G$ obtained from greedy modularity} \\
		\For{$g^{(1)} \in \mathcal{S}^{(1)}$}
		{
			$S \gets S \cup \RecLouvain(g^{(1)})$ 
		}
		\Return $\mathcal{S}$
	}
	
	\BlankLine
	
	\Block({$\RecLouvain(g^{(1)})$}) 
	{	
		$\mathcal{S}^{(3)}\gets\{\}$ \\
		$\mathcal{S}^{(2)} \gets $ \expr{a set of subgraphs of $g^{(1)}$ obtained from the Louvain method} \\
		\For{$g^{(2)} \in \mathcal{S}^{(2)} $}{
		
			\% \textit{Introduce edge nodes} \\
			$\mathcal{W} \gets []$ \\
			$\mathcal{V} \gets$ \expr{a set of nodes in $g^{(2)}$} \\
			\For{$i \in \mathcal{V}$}{
				 $\mathcal{E}_i \gets$ \expr{a set of edges in $G$ that are incident to $i$} \\
				 \For{$(i,i')\in \mathcal{E}_i$}{
					 \If{$i'\notin \mathcal{V}$ and $i'\notin \mathcal{W}$}{
						 $\mathcal{W}\gets \mathcal{W}\cup\{i'\}$
					 }
				 }
			}
			$g^{(2)} \gets$ \expr{add nodes in $\mathcal{W}$ and corresponding edges to $g^{(2)}$} \\
			\% \textit{Recursive division} \\
			\If{$|g^{(2)}|>\nu$}{
				$\mathcal{S}^{(3)}\gets \mathcal{S}^{(3)}\cup\RecLouvain(g^{(2)})$ \\
			} 
			\Else{
				$\mathcal{S}^{(3)}\gets \mathcal{S}^{(3)}\cup\{g^{(2)}\}$
			}
		}
		\Return $\mathcal{S}^{(3)}$
	}
	\caption{Recursive Division}
	\label{ag:division}
	\end{algorithm}

\begin{algorithm}[h]
\SetKwBlock{Block}{function}{}
\SetKwInOut{Input}{input}
\SetKwInOut{Output}{output}
\SetKw{Return}{return}
\SetKwFunction{Energy}{Energy}
\SetKwFunction{Evolve}{Evolve}
\Input
{(i) Local fields \{$h_{i,j}\}$ for subgraph $g$. \newline
 (ii) Association  $\{A_{i,i'}\}$ for subgraph $g$. \newline 
 (iii) The number of driving cycles $T$ in QAOA. \newline 
 (iv) The maximum number of iterations $\eta$ in the COBYLA.}

\Output{A set of probabilities $\{p(\underline{x})|\forall \underline{x} \in \{0,1\}^{N_g\times M}\}$, where $N_g$ is the number of loanees in subgroup $g$. Configurations $\underline{x}$ that have high values of the objective function will have high $p(\underline{x})$.}
\Begin
{
    $\underline{\theta}\gets$ \expr{a vector of size $2T$ containing randomized real numbers $\in [0,1)$} \\
    $\underline{\theta}_{\rm opt}\gets$ \expr{minimizes $\Energy(\underline{\theta})$ using COBYLA} \\
    $|\psi_{\rm opt}\rangle\gets\Evolve(|\psi_0\rangle, \underline{\theta}_{\rm opt})$ \\
    \Return $\{p(\underline{x})=\psi_{\rm opt}\rangle |^2\}$
}
\BlankLine
\Block({$\Energy(\underline{\theta})$}) 
{	
	$|\psi\rangle\gets\Evolve(|\psi_0\rangle, \underline{\theta})$ \\
	\Return $\langle \psi | \hat{H}_B|\psi\rangle$ 
}

\BlankLine
\Block({$\Evolve(|\psi\rangle, \underline{\theta})$}) 
{	
	\For{$t=0$ to $T-1$}{
		\% Exploitation  \\
		$\hat{U}_A\gets \exp\left(-i\hat{H}_A \cdot\underline{\theta}[2t]\right)$ \\
		\% Exploration  \\
		$\hat{U}_B\gets \exp\left(-i\hat{H}_B \cdot\underline{\theta}[2t+1]\right)$ \\
		$|\psi\rangle\gets \hat{U}_A\hat{U}_B|\psi\rangle$
	}
	\Return $|\psi\rangle$
}
\caption{QAOA with local constraints}
\label{ag:qaoa}
\end{algorithm}

\begin{algorithm}[h]
\SetKwBlock{Block}{function}{}
\SetKwInOut{Input}{input}
\SetKwInOut{Output}{output}
\SetKw{Return}{return}
\SetKwFunction{Combine}{Combine}
\Input
{(i) A set of the subgraphs $\mathcal{S}$ obtained from the Recursive Division algorithm. \newline 
(ii) A set of probabilities for every subgraphs $\{p_g(\underline{x})|\forall \underline{x} \in \{0,1\}^{N_g\times M}, \forall g\in \mathcal{S}\}$ obtained from QAOA. \newline 
(iii) A hyper-parameter for the maximum number of candidates per subgraph $\lambda$.}

\Output{A set of actions that maximises $Y$.}
\Begin
{
	$g_L\gets$ \expr{a randomly chosen subgraph from $\mathcal{S}$} \\
	
	$\mathcal{X}_{L}\gets$ \expr{a set of the first $\lambda$ bit-strings $\{\underline{x}\}$ from subgraph $g_L$ that have the highest probabilities $\{p_{gL}(\underline{x})\}$} \\
	
	\For{$g_R\in \mathcal{S} \cap \{g_L\}$}{
	
		$\mathcal{X}_{R}\gets$ \expr{a set of the first $\lambda$ bit-strings $\{\underline{x}\}$ from subgraph $g_R$ that have the highest probabilities $\{p_{g_R}(\underline{x})\}$} \\
		
		$\mathcal{W}_{LR}\gets$ \expr{a set of edge nodes between $g_{L}$ and $g_R$} \\

		$\mathcal{X}_{LR}=\{\}$ \\
		\For {$\underline{x}_L \in \mathcal{X}_L$}{
			\For {$\underline{x}_R \in \mathcal{X}_R$}{
				$\mathcal{X}_{LR} \gets \mathcal{X}_{LR}\cup\{\Combine(\underline{x}_L,\underline{x}_R, \mathcal{W}_{LR})\}$				
			}
		}

		\While {$|\mathcal{X}_{LR}|==0$}{
			$\underline{x}'_R\gets$ \expr{the next-highest-probability bit-string from $g_R$} \\
			
			\For{$\underline{x}_L\in \mathcal{X}_L$}{
				$\mathcal{X}_{LR} \gets \mathcal{X}_{LR}\cup\{\Combine(\underline{x}_L,\underline{x}'_R, \mathcal{W}_{LR})\}$	
			}
		}

		\If {$|\mathcal{X}_{LR}|>\lambda$}{
			$\mathcal{X}_{LR}\gets$ \expr{only keep the first $\lambda$ elements that have the highest $Y$}\\
		}
		
		$g_L\gets$ \expr{combine nodes and edges in $g_L$ and $g_R$.}

		$\mathcal{X}_L\gets \mathcal{X}_{LR}$\\
	}
    \Return $\mathcal{X}_L$
}
\BlankLine
\Block({$\Combine(\underline{x}_L,\underline{x}_R, \mathcal{W}_{LR})$}) 
{	
	\If{actions at the edge nodes from $\underline{x}_L$ and $\underline{x}_R$ are the same}{
		\Return \expr{a combined bit-string $\underline{x}_{LR}$ as explained in the main text.} 
	}
}
\caption{State Reconstruction}
\label{ag:reconstruction}
\end{algorithm}

\begin{algorithm}

\SetFuncSty{textsc}
\SetDataSty{emph}
\SetCommentSty{commentsty}
\SetKwComment{Comment}{$\triangleright$ }{}
\SetKwBlock{Block}{function}{}
\SetKwInOut{Input}{input}
\SetKwInOut{Output}{output}
\SetKw{Return}{return}
\SetKwData{Node}{node}
\SetKwData{Leaf}{leaf}

\Input
{(i) A set of actions $\underline{j}$, where the element $\underline{j}[i]\in \{1,2,..,M\}$ is the optimal action $j$ that is taken to loanee $i$.\newline 
(ii) The maximum number of iteration $\eta$.}

\Output{A new optimal actions that minimises the LLP.}

\Begin
{
	$\underline{j}_{\rm best}\gets \underline{j}$ \\
	$y_{\rm best}\gets$ \expr{The objective function for $\underline{j}$} \\
	$l_{\rm best}\gets$ \expr{The LLP of $\underline{j}$} \\
    \For{$k=1$ to $k=\eta$}{ 
    	$F\gets \{\}$ \\
    	\For{$i=1$ to $i=N$}{
    		\For{$j'=1$ to $j'=M$}{

    		$\underline{j}_{\rm new}\gets$\expr{a new set of actions where $\underline{j}[i]$ is changed to $j'$} \\
    		$y_{\rm new}\gets$\expr{the objective function for $\underline{j}_{\rm new}$} \\
    		$l_{\rm new}\gets$\expr{the LLP of $\underline{j}_{\rm new}$} \\
    		\% \textit{Reward} \\
    		$a_{i,j\to j'}\gets l_{\rm best} - l_{\rm new}$ \\
    		\% \textit{Penalty} \\
    		$b_{i,j\to j'}\gets y_{\rm best} - y_{\rm new}$ \\
    	    \% \textit{Finesse score} \\
    			\If{$a_{i,j\to j'} > 0$ and $b_{i,j\to j'}\le 0$}{
    				$f_{i,j\to j'}\gets a_{i,j\to j'}$ \\
    			}
    			\If{$a_{i,j\to j'} > 0$ and $b_{i,j\to j'}>0$}{
    				$f_{i,j\to j'}\gets -b_{i,j\to j'}/a_{i,j\to j'}$ \\
    			}    			
    			\If{$a_{i,j\to j'} < 0$}{
    				$f_{i,j\to j'}\gets -\infty$ \\
    			}
    			\If{$a_{i,j\to j'} = 0$ and $b_{i,j\to j'}\> 0$}{
    				$f_{i,j\to j'}\gets -\infty$ \\
    			}   
    			$F\gets F\cup\{f_{i,j\to j'}\}$
    		}	 	 			    			
		}
		$\underline{j}_{\rm best}\gets$\expr{a new set of actions where $\underline{j}[i]$ is changed to $j'$ such that $f_{i,j\to j'}\in F$ is maximized} \\
		$y_{\rm best}\gets$ \expr{The objective function for $\underline{j}$} \\
		$l_{\rm best}\gets$ \expr{The LLP for $\underline{j}$}		 \\
    }
    \Return $\underline{u}_{\rm best}$
}
\caption{Greedy Provision Reduction}
\label{aq:sampling}
\end{algorithm}

\end{widetext}

\end{document}